\documentclass[aip,jcp,reprint]{revtex4-1}

\usepackage{bm}
\usepackage{graphicx}
\usepackage{amsmath}
\usepackage{amssymb}
\usepackage{array}

\providecommand{\eqnref}[1]{Eq.\ \eqref{#1}}
\providecommand{\figref}[1]{Fig.\ \ref{#1}}
\providecommand{\tabref}[1]{Table \ref{#1}}
\providecommand{\mat}[1]{\ensuremath{\bm{\mathsf{#1}}}}
\providecommand{\trans}[1]{\ensuremath{#1^{\intercal}}}
\providecommand{\dtmt}[1]{\ensuremath{|#1|}}
\renewcommand{\vec}[1]{\ensuremath{\mathbf{#1}}}
\DeclareMathOperator{\adj}{adj}
\DeclareMathOperator{\tr}{Tr}

\begin{document}
\title{Dirac cones in the spectrum of bond-decorated graphenes}
\author{Willem Van den Heuvel}
\email{wvan@unimelb.edu.au}
\author{Alessandro Soncini}
\email{asoncini@unimelb.edu.au}
\affiliation{School of Chemistry, The University of Melbourne, VIC 3010,
Australia}
\date{\today}

\begin{abstract} We present a two-band model based on periodic H\"uckel theory,
which is capable of predicting the existence and position of Dirac cones in the
first Brillouin zone of an infinite class of two-dimensional periodic carbon
networks, obtained by systematic perturbation of the graphene connectivity by
\emph{bond decoration}, that is by inclusion of arbitrary $\pi$-electron
H\"uckel networks into each of the three carbon--carbon $\pi$-bonds within the
graphene unit cell.  The bond decoration process can fundamentally modify the
graphene unit cell and honeycomb connectivity, representing a simple and
general way to describe many cases of graphene chemical functionalization of
experimental interest, such as graphyne, janusgraphenes and chlorographenes.
Exact mathematical conditions for the presence of Dirac cones in the spectrum
of the resulting two-dimensional $\pi$-networks are formulated in terms of the
spectral properties of the decorating graphs.  Our method predicts the
existence of Dirac cones in experimentally characterized janusgraphenes and
chlorographenes, recently speculated on the basis of DFT calculations.  For
these cases, our approach provides a proof of the existence of Dirac cones, and
can be carried out at the cost of a back of the envelope calculation, bypassing
any diagonalization step, even within H\"uckel theory.  
\end{abstract}

\maketitle

\section{Introduction}

A peculiar feature in the band structure of graphene is the touching of valence
and conduction bands at the corners of the Brillouin zone, and the linear and
isotropic dispersion in the neighborhood of these points, giving rise to two
characteristic valence and conduction conical features with touching vertices
at the Fermi energy, also known as Dirac cones. Electronic excitations close to
the Fermi energy in graphene can be described via an effective low-energy
theory based on a relativistic fermionic Hamiltonian.\cite{CastroNeto2009} This
pseudo-relativistic model unveiled the remarkable properties of low-energy
electronic excitations in graphene, which behave like Dirac fermions, and give
rise to new physical phenomena previously unobserved in condensed matter, such
as the anomalous half-integer quantum Hall effect\cite{Novoselov2005} and the
Klein paradox.\cite{Katsnelson2006}  Most of these effects have been
experimentally confirmed, and offer promising strategies for the development of
a future generation of graphene-based electronic devices.

The necessary condition to observe this pseudo-relativistic electronic behavior
is the existence of Dirac cones in the spectrum of a two-dimensional
carbon network, which, quite remarkably, are correctly predicted already at the
simple H\"uckel level of theory in the case of the graphene honeycomb lattice.
One of the powerful features of H\"uckel theory is that it provides an
intuitive link between a molecule's connectivity and its electronic structure,
and chemical and physical properties.  Thus, from a chemist's point of view, an
interesting question in connection to the chemical design of two-dimensional
systems displaying Dirac cone physics is: How is the existence of Dirac cones
related to the details of the sp$^2$-carbon connectivity, and its departure
from the graphene honeycomb structure?  While it is in fact known that Dirac
cones appear in the spectrum of other two-dimensional systems preserving to
some degree the graphene honeycomb connectivity,  as for instance in some
isomers of graphyne \cite{Kim2012,Zheng2013} and some graphene antidot
lattices,\cite{Ouyang2011} in recent reports on periodic covalent functionalization of
graphene resulting in janusgraphenes and chlorographenes, where the departure
of the resulting sp$^2$-carbon network from the honeycomb connectivity is
significant,\cite{Yang2012,Zhang2013,Ma2013} it has been proposed on the basis
of DFT calculations that putative Dirac cones are retained in the band
structure of the functionalized graphenes.

In this work we address this question using periodic H\"uckel theory (also
known as orthogonal tight binding), by constructing a class of two-dimensional
H\"uckel graphs, which are derived from the graphene honeycomb lattice by bond
decoration, i.e.\ by insertion of arbitrary H\"uckel graph fragments (or
H\"uckel molecules) in each of the three translationally unique carbon--carbon
bonds in the unit cell of graphene. This is a procedure that has previously
been applied to planar conjugated molecules in order to explain their
electronic structure and magnetic properties in terms of those of the parent
carbocycles from which they could be derived upon bond decoration or atomic
decoration.\cite{Soncini2001,Soncini2005a,Soncini2005}  In these previous
works, it was shown that perturbation of the $\pi$-electron connectivity via
bond decoration, as opposed to simple structural distortion, can radically
change the electronic structure of the resulting $\pi$ network, and, as a
consequence, its response to external magnetic
fields.\cite{Soncini2001,Soncini2005a,Soncini2005,Fowler2001,Soncini2002} 
Likewise the general class of bond-decorating perturbations to the graphene's
$\pi$ connectivity explored here will lead to a very diverse set of
two-dimensional structures, where the honeycomb connectivity of the parent
graphene graph is typically lost, as will be, in general, the Dirac cones in
their band structure. We examine the conditions for the presence of Dirac
cones in this particular class of systems, reducing the problem to the analysis
of three matrix elements in an effective two-band model.  It will be shown that
the proposed method provides a straightforward strategy to screen
two-dimensional carbon networks for the existence of Dirac cones, avoiding
costly all-electron approaches such as DFT, or even full diagonalization of the
bond-decorated H\"uckel problem.  For a few cases of experimental interest,
such as the janusgraphenes and chlorographenes recently synthesized and
discussed in the literature, our method offers a full analytical tool to prove
the existence of Dirac cones, where numerical ab initio or DFT approaches can
in principle only suggest the existence of an exactly degenerate Dirac point
within the numerical precision of the calculation.

\section{H\"uckel theory of bond-decorated graphene}

\subsection{Definition of the decorated graphene lattice}
\begin{figure}
\centering
\includegraphics[width=8.5cm]{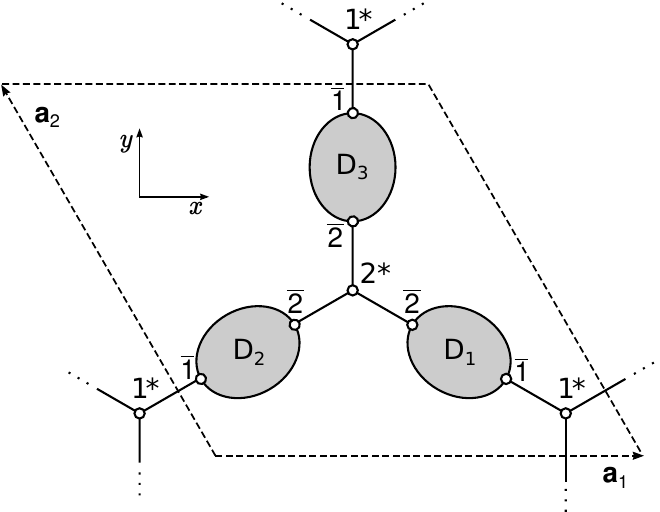}
\caption{Primitive unit cell of bond-decorated graphene. $\vec{a}_1=(1,0)$,
$\vec{a}_2=(-1/2,\sqrt{3}/2)$.}
\label{fig:dec}
\end{figure}

The systems we shall be discussing are derived from the honeycomb lattice, whose
primitive unit cell contains two vertices and three edges. The honeycomb
vertices are marked with a star, as 1$^\ast$ and 2$^\ast$, to distinguish them
from vertices that belong to the decorations.
We insert in each of the three edges an arbitrary ``molecular'' graph (the
\emph{decoration}), which shall be connected to the starred vertices by a
single edge. There are thus at most three different decorating graphs, one for
each original edge in the unit cell. For each decoration, the vertex that is
connected to 1$^\ast$ shall be labeled $\bar{1}$ and the vertex that is
connected to 2$^\ast$ shall be labeled $\bar{2}$. Every unit cell of the
honeycomb lattice is decorated in the same way. This construction, which is
illustrated in \figref{fig:dec}, ensures that the starred vertices retain the
trivalent sp$^2$ valency of the honeycomb graph and that the periodicity of the
latter is conserved in the starred sublattice.

\subsection{Bloch Hamiltonian}
We describe the band structure by a H\"uckel Hamiltonian with on-site Coulomb
energy $\alpha=0$ and transfer integral $\beta=1$. In this conventional setting
the Hamiltonian is given by the adjacency matrix of the graph. Upon forming
Bloch waves of the vertices the adjacency matrix factors into blocks labeled by
wavevector \vec{k}:
\begin{equation}\label{eq:bloch}
\mat{H}(\vec{k})= 
\begin{pmatrix}
0 & \gamma(\vec{k}) & \vec{e}_{11} & e^{i\vec{k}\cdot\vec{a}_1}\vec{e}_{12}& 
 e^{-i\vec{k}\cdot\vec{a}_2}\vec{e}_{13}\\
\overline{\gamma}(\vec{k}) & 0 & \vec{e}_{21}& \vec{e}_{22}& \vec{e}_{23}\\
\trans{\vec{e}}_{11} & \trans{\vec{e}}_{21} & \mat{D}_1 & \mat{0} & \mat{0}\\
e^{-i\vec{k}\cdot\vec{a}_1}\trans{\vec{e}}_{12} & \trans{\vec{e}}_{22} &
\mat{0} &\mat{D}_2 & \mat{0} \\
e^{i\vec{k}\cdot\vec{a}_2}\trans{\vec{e}}_{13} & \trans{\vec{e}}_{23} & \mat{0}
& \mat{0}& \mat{D}_3
\end{pmatrix}
\end{equation}
Here $\mat{D}_\kappa$ are the adjacency matrices of the decorating fragments,
and $\vec{e}_{1\kappa}$ and $\vec{e}_{2\kappa}$ are row vectors indicating
which vertices of decoration $\kappa$ connect to the honeycomb vertices
1$^\ast$ and 2$^\ast$, respectively: $[\vec{e}_{1\kappa}]_i=\delta_{i\bar{1}}$
and $[\vec{e}_{2\kappa}]_i=\delta_{i\bar{2}}$. In expression \eqref{eq:bloch}
the first two rows/columns represent the vertices 1$^\ast$ and 2$^\ast$. The
interaction element $\gamma$ can be expressed as 
\begin{equation*}
\gamma(\vec{k})=\delta_1+\delta_2e^{i\vec{k}\cdot\vec{a}_1}+\delta_3
e^{-i\vec{k}\cdot\vec{a}_2},
\end{equation*}
where $\delta_\kappa=1$ if bond $\kappa$ is \emph{not} decorated (i.e., if a
direct edge connects 1$^\ast$ and 2$^\ast$ along bond $\kappa$), and 0
otherwise. Note that, in case bond $\kappa$ is not decorated, the columns and
rows corresponding to $\mat{D}_\kappa$ should be removed from \mat{H}.  

\subsection{Effective two-band Hamiltonian}
As we are interested in the presence of Dirac cones in the band spectrum, we
shall be looking for discrete doubly degenerate zero-energy eigenvalues of the
Bloch Hamiltonian $\mat{H}(\vec{k})$. The approach followed
here is to eliminate the bond decorations from the Hamiltonian, to obtain a
low-energy two-band Hamiltonian on the honeycomb lattice with effective Coulomb
and transfer integrals. 

We write $\mat{H}(\vec{k})$ in a compact notation as
\begin{equation}\label{eq:Hpartition}
\mat{H}= 
\begin{pmatrix}
\mat{H}^\ast & \mat{T} \\
\mat{T}^\dagger & \mat{H}_{\mat{D}}
\end{pmatrix}
\end{equation}
and use the Schur complement formula (Ref.\
\onlinecite{Horn1990}, 0.8.5, p.\ 21) to obtain a L\"owdin partitioning of the secular
determinant:
\begin{equation}\label{eq:schur}
P(\vec{k},E)\equiv\dtmt{E-\mat{H}}= \dtmt{E-\mat{H}_{\mat{D}}}
\dtmt{E-\mat{H}^\ast-\mat{T}(E-\mat{H}_{\mat{D}})^{-1}\mat{T}^\dagger}.
\end{equation}
Introducing now the effective $2\times 2$ Hamiltonian on the honeycomb lattice
\begin{equation}\label{eq:effH}
\widetilde{\mat{H}^\ast}(E)= \mat{H}^\ast+\mat{T}(E-\mat{H}_{\mat{D}})^{-1}\mat{T}^\dagger
\end{equation}
and expanding $\mat{H}_{\mat{D}}$ gives
\begin{equation}\label{eq:charpol}
P(\vec{k},E)=
\dtmt{E-\mat{D}_1}\dtmt{E-\mat{D}_2}\dtmt{E-\mat{D}_3}
\dtmt{E-\widetilde{\mat{H}^\ast}(E)}.
\end{equation}
Zero-energy eigenvalues exist if the equation $P(\vec{k},0)=0$ has a solution. A
problem arises when 0 is a root of one or more of the decorating units
$\mat{D}_\kappa$, for then the inverse of $\mat{H}_{\mat{D}}$ does not exist
and the effective Hamiltonian is not defined at $E=0$, as can be seen from
\eqnref{eq:effH}. However, as we will show below in Sec.~\ref{sec:deczeros}, this
problem can be analyzed by factoring the dispersionless zero-energy poles out
of each decoration's Green's function.

Let us now represent the effective Hamiltonian as follows, indicating explicitly
its dependence on \vec{k} and $E$ :
\begin{equation}\label{eq:effHmat}
\widetilde{\mat{H}^\ast}(\vec{k},E)=
\begin{pmatrix}
\alpha_1(E) & \Gamma(\vec{k},E)\\
\overline{\Gamma}(\vec{k},E) & \alpha_2(E)
\end{pmatrix}
\end{equation}
Define Green's functions for the decorations:
\begin{equation}\label{eq:decogf}
\mat{G}^\kappa(E)=
(E-\mat{D}_\kappa)^{-1}=\frac{1}{\dtmt{E-\mat{D}_\kappa}}\adj
(E-\mat{D}_\kappa).
\end{equation}
Then we find\footnote{Assuming for now that all three bonds are decorated, so that
$\gamma=0$. This is done for generality only; it does not represent a limitation of
the theory. The example in Sec.\ \ref{sec:example} will show how to treat
undecorated bonds on the same footing as decorated bonds.}
\begin{equation}\label{eq:effme}
\begin{split}
\alpha_1(E)&=\mat{G}^1_{\bar{1}\bar{1}}(E) + \mat{G}^2_{\bar{1}\bar{1}}(E)
+\mat{G}^3_{\bar{1}\bar{1}}(E) \\
\alpha_2(E)&=\mat{G}^1_{\bar{2}\bar{2}}(E) + \mat{G}^2_{\bar{2}\bar{2}}(E)
+\mat{G}^3_{\bar{2}\bar{2}} (E)\\
\Gamma(\vec{k},E)&= \mat{G}^1_{\bar{1}\bar{2}}(E) +e^{i\vec{k}\cdot\vec{a}_1}
\mat{G}^2_{\bar{1}\bar{2}}(E)+e^{-i\vec{k}\cdot\vec{a}_2}\mat{G}^3_{\bar{1}\bar{2}}(E)
\end{split}
\end{equation}
The matrix elements in this expression follow from evaluation of
\eqnref{eq:effH}:
$\mat{G}^\kappa_{\bar{i}\bar{j}}=\vec{e}_{i\kappa}\cdot\mat{G}^\kappa\cdot\trans{\vec{e}_{j\kappa}}$.
They represent the contribution from the decorations to the Coulomb integral on
($\alpha_1$ and $\alpha_2$) and the transfer/resonance integral between
($\Gamma$) the starred vertices of the honeycomb lattice.

Inspection of expression \eqref{eq:effHmat} learns that a pair of zero-energy
solutions at wavevector \vec{k} exists if and only if
$\alpha_1(0)=\alpha_2(0)=0$ and $\Gamma(\vec{k},0)=0$.  

\subsubsection{Decorations with zero eigenvalues}\label{sec:deczeros}

Expressions \eqnref{eq:decogf} and \eqnref{eq:effme} become ill-defined if the
H\"uckel adjacency matrix $\mat{D}_\kappa$ for one or more of the decorating
graphs has zero eigenvalues.  This case can be dealt with in some detail if the
non-zero eigenvalues of the decorating graphs are known.  For simplicity we
discuss here the case of uniform decoration, where all three carbon--carbon
bonds in the unit cell of graphene are decorated by the same chemical graph
with $N$ vertices, having an adjacency matrix $\mat{D}$ with $n<N$ zero
eigenvalues (i.e.\ the decoration's adjacency matrix has rank $m=N-n$) and $m$
non-zero eigenvalues $\lambda_1, \dots, \lambda_{m}$.  The analysis can be
generalized to an arbitrary set of decorating graphs with an arbitrary number
of zero eigenvalues, in a similar manner. We start off by writing an explicit
expression for the decoration secular determinant in terms of its eigenvalues:
\begin{equation}\label{eq:decodet1}
\dtmt{E-\mat{D}_\kappa}\equiv \dtmt{E-\mat{D}}  = E^{n}
\dtmt{E-\mat{\tilde{D}}},
\end{equation}
where $\dtmt{E-\mat{\tilde{D}}}$ is defined solely in terms of the $m$ non-zero
eigenvalues as: 
\begin{equation}\label{eq:decodet2}
\dtmt{E-\mat{\tilde{D}}} = (E-\lambda_1)(E-\lambda_2)\dots (E-\lambda_{m}).
\end{equation}
Using \eqnref{eq:decodet1} to express the denominators in \eqnref{eq:decogf}
and \eqnref{eq:effme}, the effective two-band Hamiltonian can now be written in
a form that explicitly display its poles at $E=0$:
\begin{equation}\label{eq:effHmatPoles}
\begin{split}
\widetilde{\mat{H}^\ast}(\vec{k},E)&
= 
\frac{1}{E^{n}} \sum_{\kappa =1,2,3}  \mat{e}_\kappa(\vec{k}) \cdot 
\frac{\adj(E-\mat{D})} {\dtmt{E-\mat{\tilde{D}}} } 
\cdot \trans{\mat{e}_\kappa}(\vec{k})\\
&\equiv \frac{1}{E^{n}} \mat{M}(\vec{k},E)
\end{split}
\end{equation}

If we label the two eigenvalues of the pole-free effective two-band Hamiltonian
$\mat{M}(E,\vec{k})$ introduced in~\eqnref{eq:effHmatPoles} as
$\eta_1(\vec{k},E)$ and  $\eta_2(\vec{k},E)$, the secular polynomial for the
decorated graphene can now be written as 
\begin{multline}\label{eq:charpol2}
P(\vec{k},E)= \dtmt{E-\mat{\tilde{D}}}^3 \times E^{3n} \\\times
(E-\frac{\eta_1(\vec{k},E) }{E^{n}} )\times(E-\frac{\eta_2(\vec{k},E) }{E^{n}}
).  
\end{multline} From~\eqnref{eq:charpol2} we can draw two conclusions.

First, as $E\rightarrow0$, the denominator tends to zero as $E^{2n}$, while the
numerator tends to zero as $E^{3n}$, which leads to a pole-free secular
polynomial having a flat zero-energy band which is at least $n$-fold degenerate
(in fact, for $n>1$, it can be seen that additional degeneracy of the flat
zero-energy band arises from $\adj(E-\mat{D})$).

Second and most important, by factoring out the poles of the singular effective
Hamiltonian $\widetilde{\mat{H}^\ast}(\vec{k},E)$ in this manner, we arrive at
a reformulation of the secular polynomial of bond-decorated graphene where the
problem of finding Dirac cones is reduced to the problem of finding the
zero-energy solutions of a new non-singular effective two-band Hamiltonian
$\mat{M}(\vec{k},E)$ (see~\eqnref{eq:effHmatPoles}), and is thus confined to
the dispersion structure of the two eigenvalues $\eta_1(\vec{k},E)$ and
$\eta_2(\vec{k},E)$.  Necessary conditions for the existence of Dirac points
can thus be derived in the same manner as for the case of bond-decorating
graphs where the adjacency-rank is equal to the number of vertices (see
\eqnref{eq:effme} and the following discussion).

\subsection{Criteria for the existence of Dirac points}

In order to simplify expressions in the discussions that follow we introduce 
the notation of Pickup and Fowler\cite{Pickup2008} for the polynomials that
appear in Eqs.\ \eqref{eq:decogf} and \eqref{eq:effme}:
\begin{equation}\label{eq:polyn}
\begin{split}
s_\kappa(E) &= \dtmt{E-\mat{D}_\kappa}\\
t_\kappa(E) &= \dtmt{(E-\mat{D}_\kappa)^{\bar{1},\bar{1}}}\\
u_\kappa(E) &= \dtmt{(E-\mat{D}_\kappa)^{\bar{2},\bar{2}}}\\
v_\kappa(E) &= \dtmt{(E-\mat{D}_\kappa)^{\bar{1}\bar{2},\bar{1}\bar{2}}}\\
w_\kappa(E) &= (-1)^{\bar{1}+\bar{2}}\dtmt{(E-\mat{D}_\kappa)^{\bar{2},\bar{1}}},
\end{split}
\end{equation}
where $\mat{M}^{i_1i_2\ldots,j_1j_2\ldots}$ is the matrix derived from
$\mat{M}$ by removing rows $i_1,i_2,\ldots$ and columns $j_1,j_2,\ldots$. 
The polynomials in \eqnref{eq:polyn} are related by the Sylvester identity for
determinants (See Appendix 2 in Ref.\ \onlinecite{Gutman1986}), which reads
\begin{equation}\label{eq:sylv}
t_\kappa u_\kappa-s_\kappa v_\kappa=w_\kappa^2.
\end{equation}
\eqnref{eq:effme} can now be written as
\begin{equation}\label{eq:mepoly}
\begin{split}
\alpha_1&=\frac{t_1}{s_1}+\frac{t_2}{s_2}+\frac{t_3}{s_3}\\
\alpha_2&=\frac{u_1}{s_1}+\frac{u_2}{s_2}+\frac{u_3}{s_3}\\
\Gamma&= \frac{w_1}{s_1}+
e^{i\theta_1}\frac{w_2}{s_2}+e^{-i\theta_2}\frac{w_3}{s_3},
\end{split}
\end{equation}
where $\theta_i=\vec{k}\cdot\vec{a}_i$. 

A spectrum with Dirac cones is characterized by a discrete number of wavevectors at
which two bands cross linearly with energy equal to zero (this is the Fermi
energy). These wavevectors are called Dirac points in the Brillouin zone. For
simplicity we restrict attention to decorations that have no zero eigenvalues. 
This means that $s_\kappa\neq0$ at $E=0$ for all $\kappa$. Then, as
seen before, Dirac points are solutions of the equations
$\alpha_1(0)=\alpha_2(0)=0$ and $\Gamma(\vec{k},0)=0$. Note that only the last
equation determines, implicitly, the position of the Dirac points, if any. The
first two equations are wavevector-independent. They read, according to
\eqnref{eq:mepoly},
\begin{equation}\label{eq:dpeq1}
\begin{split}
\frac{t_1}{s_1}+\frac{t_2}{s_2}+\frac{t_3}{s_3}&=0 \quad \text{at } E=0\\
\frac{u_1}{s_1}+\frac{u_2}{s_2}+\frac{u_3}{s_3}&=0\quad \text{at } E=0.
\end{split}
\end{equation}
The equations can be satisfied by an accidental cancellation of terms, but more
relevant are cases in which the terms vanish individually, i.e.\
$t_1=t_2=t_3=0$ at $E=0$, and $u_1=u_2=u_3=0$ at $E=0$. In the rest of the
paper we shall be considering only this case.

In terms of graph theory we know that $s_\kappa$ is the characteristic
polynomial of the graph of decoration $\kappa$. From \eqnref{eq:polyn}
$t_\kappa$ is seen to be the characteristic polynomial of a subgraph of
$\kappa$, namely the subgraph obtained by removing vertex $\bar{1}$, which is
the one that connects to vertex $1^\ast$ of the honeycomb
lattice.\cite{Pickup2008} Similarly, $u_\kappa$ is the characteristic
polynomial of the graph $\kappa$ with vertex $\bar{2}$ removed. Note that
$\bar{1}$ and $\bar{2}$ can be the same vertex. In that case
$u_\kappa=t_\kappa$. 

From the preceding discussion we conclude that the first condition for the
presence of Dirac points, expressed in \eqnref{eq:dpeq1}, is a condition on
zeros in the spectrum of the decorating graphs and subgraphs. This condition
states that none of the three decorating graphs can have nonbonding orbitals
while each of the six single-vertex-deleted subgraphs must have a nonbonding
orbital. 

The second condition states that the off-diagonal element in the effective
two-band Hamiltonian \eqref{eq:effHmat} is zero at the Fermi energy $E=0$. From
\eqnref{eq:mepoly}, this means
\begin{equation*}
\Gamma(\vec{k},0)\equiv\frac{w_1}{s_1}+ e^{i\theta_1}\frac{w_2}{s_2}+e^{-i\theta_2}\frac{w_3}{s_3}=0
\quad \text{at } E=0.
\end{equation*}
We note that $\Gamma(\vec{k},0)$ can be seen as the off-diagonal element in the
two-band tight binding model of the anisotropic honeycomb lattice
\cite{Hasegawa2006} if we identify the hopping integrals $\beta_\kappa$ of
the latter with the polynomial ratios in our expression: $\beta_\kappa\equiv
\frac{w_\kappa}{s_\kappa}|_{E=0}$. The off-diagonal condition can then be
written as
\begin{equation}\label{eq:triangle}
\beta_1+e^{i\theta_1}\beta_2+e^{-i\theta_2}\beta_3=0.
\end{equation}
Solving this equation for $(\theta_1,\theta_2)$ gives the positions in the
Brillouin zone where two bands are degenerate at $E=0$ (assuming that the
conditions \eqnref{eq:dpeq1} on the diagonal elements are also fulfilled).
Given that the $\beta_\kappa$ are all real, \eqnref{eq:triangle} describes a
triangle in the complex plane, with sides of length $|\beta_1|$, $|\beta_2|$,
and $|\beta_3|$. Hence a solution to \eqnref{eq:triangle} exists only when the
effective hopping parameters satisfy the triangle inequality
\cite{Hasegawa2006,Kim2012}
\begin{equation}\label{eq:trianineq}
(|\beta_1|-|\beta_2|)^2 \leq |\beta_3|^2 \leq (|\beta_1|+|\beta_2|)^2.
\end{equation}
In order to have a discrete number of Dirac points we must also require that
none of the $\beta_\kappa$ be zero. Consider the two possibilities: (i)
$\beta_1=\beta_2=\beta_3=0$. Then \eqnref{eq:triangle} is automatically
satisfied over the entire Brillouin zone, which means that there are two flat
bands at $E=0$ but no Dirac cones; (ii) $\beta_1=0$ and $|\beta_2|=|\beta_3|$,
or a circular permutation of this. Then solution of \eqnref{eq:triangle} gives
a one-dimensional domain in the Brillouin zone (namely, straight line(s)), in
which two bands are degenerate at $E=0$. In both cases (i) and (ii) isolated
Dirac points do not exist. If, then, none of the $\beta_\kappa$ is zero and the
triangle inequality \eqref{eq:trianineq} is satisfied, there are just two
solutions to \eqnref{eq:triangle} and hence two Dirac points. These solutions
are related by time reversal: $(\theta_1,\theta_2) \leftrightarrow
(-\theta_1,-\theta_2)$. 

Let us now see what $\beta_\kappa\equiv\frac{w_\kappa}{s_\kappa}|_{E=0}\neq0$
means in terms of the spectral properties of the decoration. The definition of
$w_\kappa$ in \eqnref{eq:polyn} shows that $w_\kappa$ is not the characteristic
polynomial of some graph derived from the graph of $\kappa$. It is, however,
related to such polynomials via the Sylvester identity \eqnref{eq:sylv}.
Recall from the discussion preceding and following \eqnref{eq:dpeq1} that we
are considering decorations for which $s_\kappa\neq0$ at $E=0$ and
$t_\kappa=u_\kappa=0$ at $E=0$. Then \eqnref{eq:sylv} tells us that
$w_\kappa^2|_{E=0}=s_\kappa v_\kappa|_{E=0}$, implying that $w_\kappa$, and
thus $\beta_\kappa$, is non-zero at $E=0$ when $v_\kappa$ is non-zero at $E=0$.
$v_\kappa$ is the characteristic polynomial of the graph of decoration $\kappa$
with vertices $\bar{1}$ and $\bar{2}$ removed.\cite{Pickup2008} Hence we
conclude that $\beta_\kappa$ is non-zero iff the
$(\bar{1},\bar{2})$-vertex-deleted subgraph of decoration $\kappa$ has no
non-bonding orbitals.

In summary, discrete Dirac points in the spectrum of bond-decorated graphene
exist if all of the following statements about the spectral properties of the three
decorating graphs are true:
\begin{enumerate}
\item None of the three decorating graphs has non-bonding orbitals.
\item Each of the six subgraphs obtained by deleting the contacting vertices
$\bar{1}$ or $\bar{2}$ has at least one non-bonding orbital. 
\item None of the three subgraphs obtained by deleting both contacting vertices   
$\bar{1}$ and $\bar{2}$ has non-bonding orbitals.
\item $(|\beta_1|-|\beta_2|)^2 \leq |\beta_3|^2 \leq (|\beta_1|+|\beta_2|)^2$,
where $\beta_\kappa=\frac{w_\kappa}{s_\kappa}|_{E=0}$.
\end{enumerate}

\section{Example: Chlorinated graphene}\label{sec:example}

\begin{figure*}
\includegraphics[width=17cm]{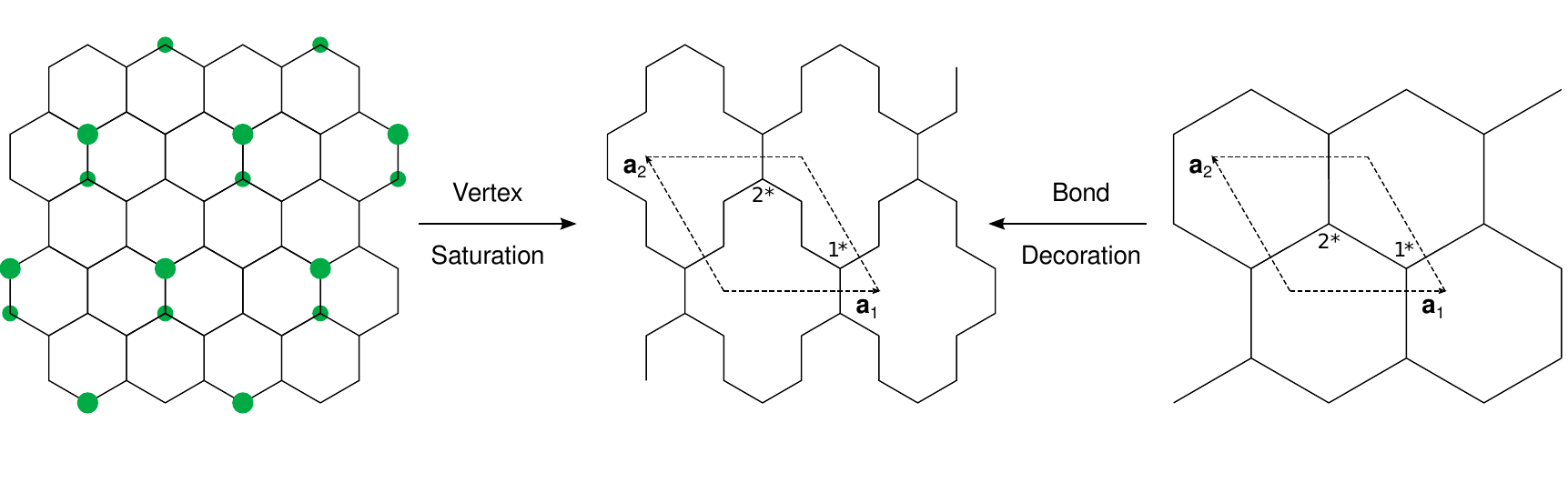}
\caption{Chlorination pattern of C$_4$Cl chlorinated graphene (left).
Chlorinated carbon atoms, which are sp$^3$ hybridized and do not contribute to the
conjugated $\pi$ electron system, are deleted to obtain the H\"uckel graph
(middle). The latter can be derived formally from graphene by decorating two
of the three bonds in the unit cell with ethylene (right $\rightarrow$ middle).}\label{fig:c4cl}
\end{figure*}

Chemically functionalized graphenes are systems where the findings of this
paper can apply. We want to illustrate this with the example of a chlorinated
graphene sheet whose stoichiometry is C$_4$Cl. The chlorine atoms in this
compound form single covalent bonds with carbon atoms of the graphene sheet.
The C--Cl bonds are arranged in a periodical pattern (i.e., translationally
symmetric) and are distributed such that half of the chlorine atoms are found
on one side of the graphene plane and the other half on the other side (see
\figref{fig:c4cl}). In a DFT study, this compound was found to be the most
stable one among double-sided chlorinated graphenes.\cite{Yang2012} Its
experimental realization was reported recently.\cite{Zhang2013}

From the point of view of electronic structure the effect of the covalent
bonding to chlorine is to change the hybridization of the carbon atom from
sp$^2$ to sp$^3$. The sp$^3$ carbon atoms do not contribute to the 
$\pi$-electron structure and may therefore be discarded. What remains is a
network of sp$^2$ carbon atoms whose band structure can be determined, in first
approximation, with H\"uckel theory. \figref{fig:c4cl} shows how the sp$^2$ network
is obtained for the C$_4$Cl example under consideration. It is apparent from
\figref{fig:c4cl} that the network can be interpreted as a decorated graphene
lattice in the sense of this paper. Two bonds are decorated with an
``ethylene'' unit. The third bond is not decorated. Hence
$\gamma(\vec{k})=e^{-i\vec{k}\cdot \vec{a}_2}$
and
\begin{equation}\label{eq:decmatrix}
\mat{D}_1=\mat{D}_2=\begin{pmatrix}
0 & 1 \\
1 & 0
\end{pmatrix}.
\end{equation}

\begin{table}
\caption{Polynomials defined in \eqnref{eq:polyn}, calculated for the C$_4$Cl
system depicted in \figref{fig:c4cl}.}\label{tb:polyn}
\begin{center}
\begin{tabular}{clllll}
\hline
Decoration & $s$ & $t$ & $u$ & $v$ & $w$\\
\hline
1 & $E^2-1$ & $E$ & $E$ & 1 & $1$ \\
2 & $E^2-1$ & $E$ & $E$ & 1 & $1$ \\
3 & $1$ & $0$ & $0$ & $-1$ & $1$ \\
\hline
& $s(0)$ & $t(0)$ & $u(0)$ & $v(0)$ & $w(0)$ \\
\hline
1 & $-1$ & $0$ & $0$ & 1 & $1$ \\
2 & $-1$ & $0$ & $0$ & 1 & $1$ \\
3 & $1$ & $0$ & $0$ & $-1$ & $1$ \\
\end{tabular}
\end{center}
\end{table}

\tabref{tb:polyn} collects the different polynomials that characterize the
decorations. For the undecorated bond (nr 3) we have chosen values for these
polynomials that are consistent with \eqnref{eq:sylv} and that can be used in
equations like \eqnref{eq:dpeq1}. This allows us to treat all three bonds on
equal footing in the discussion that follows.

To determine whether Dirac points exist we need to inspect the values of the
polynomials at $E=0$. These are given in the bottom half of \tabref{tb:polyn}.
It is straightforward to check that the first three of the four conditions
formulated at the end of Section 2 hold: For each of the three decorations,
$s(0)$ is non-zero, $t(0)$ and $u(0)$ are zero, and $v(0)$ is non-zero. To
check the fourth condition we first compute the effective hopping integrals
$\beta_\kappa=\frac{w_\kappa(0)}{s_\kappa(0)}$, giving $\beta_1=\beta_2=-1$ and
$\beta_3=1$. These satisfy the triangle inequality, and therefore the fourth
condition is also satisfied. We conclude that the H\"uckel band spectrum of CCl$_4$
has two Dirac points.

The positions of the Dirac points in the Brillouin zone are obtained from
\eqnref{eq:triangle}, which reads
\begin{equation*}
-1-e^{i\theta_1}+e^{-i\theta_2}=0.
\end{equation*}
This equation has two solutions: $(\theta_1,\theta_2)=(2\pi/3,-\pi/3)$ and
$(\theta_1,\theta_2)=(-2\pi/3,\pi/3)$. Recalling that
$\theta_i=\vec{k}\cdot\vec{a}_i$, we find that the Dirac points are located at
the wavevectors $\vec{k}=(2\pi/3,0)$ and $\vec{k}=(-2\pi/3,0)$. 

\begin{figure}
\centering
\begin{minipage}{5.5cm}
 \centering
 \includegraphics[width=5cm]{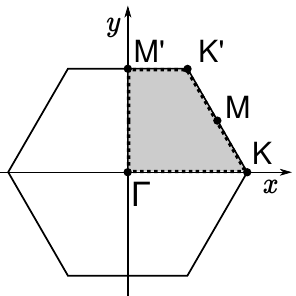}
\end{minipage}
\begin{minipage}{3cm}
\centering
\begin{tabular}{l|l}
& $(x,y)$\\
\hline
$\Gamma$ & $(0,0)$ \\
K & $(\frac{2}{3},0)$\\
M & $(\frac{1}{2},\frac{1}{2\sqrt{3}})$\\
K'& $(\frac{1}{3},\frac{1}{\sqrt{3}})$\\
M'& $(0,\frac{1}{\sqrt{3}})$
\end{tabular}
\end{minipage}
\caption{Brillouin zone for the C$_4$Cl H\"uckel system depicted in
\figref{fig:c4cl}. The symmetry-unique basic domain (grey area) is defined by
the $D_{\mathrm{2h}}$ point group symmetry of the H\"uckel graph. The dashed
line indicates the path along which the band energies are plotted in
\figref{fig:bands}. The coordinates in the table are in units of $2\pi$.}
\label{fig:BZ}
\end{figure}

\subsection{Analytical solution of the H\"uckel band spectrum}

Although the Bloch Hamiltonian for the the CCl$_4$ H\"uckel graph is a $6\times
6$ matrix, closed-form expressions for its eigenvalues can be obtained. This is
due to the matrix being bipartite, which means that its rows/columns can be
divided in two groups such that all the matrix elements that connect one group
to the other vanish. When the two groups are of equal size (as is the case for
the current example), it is known that the energies occur in pairs of opposite
sign.\cite{Coulson1940} It follows that the energies squared are the roots of a
cubic polynomial.

The Bloch Hamiltonian for the CCl$_4$ H\"uckel problem is given by
\eqnref{eq:bloch} with $\gamma(\vec{k})=e^{-i\theta_2}$, $\vec{e}_{11}=(1\;
0)=\vec{e}_{12}$, $\vec{e}_{21}=(0\; 1)=\vec{e}_{22}$, and $\mat{D}_1$ and
$\mat{D}_2$ as in \eqnref{eq:decmatrix}. The entries in \eqnref{eq:bloch} that
refer to bond 3 are omitted as this bond is not decorated. 
\begin{equation*}
\mat{H}(\vec{k})=\begin{pmatrix}
0 & e^{-i\theta_2} & 1 & 0 & e^{i\theta_1} &0 \\
e^{i\theta_2} & 0  & 0 & 1 & 0 & 1 \\
1 & 0 & 0 & 1 & 0 & 0\\
0 & 1 & 1 & 0 & 0 & 0\\
e^{-i\theta_1} & 0 & 0 & 0 & 0 & 1\\
0 & 1 & 0 & 0 & 1 & 0\\
\end{pmatrix}.
\end{equation*}
Collecting rows/columns 1, 4, and 6 in group $A$ and 2, 3, and 5 in group $B$
and rearranging the matrix accordingly clarifies its bipartite structure:
\begin{equation}\label{eq:Hbipartite}
\begin{split}
&\mat{H}(\vec{k})=\begin{pmatrix}
\mat{0} & \mat{H}_{AB}(\vec{k}) \\
\mat{H}_{AB}^\dagger(\vec{k}) & \mat{0}
\end{pmatrix}, \\
&\mat{H}_{AB}(\vec{k})=\begin{pmatrix}
e^{-i\theta_2}&1&e^{i\theta_1}\\
1&1&0\\
1&0&1\end{pmatrix}
\end{split}
\end{equation}
The characteristic polynomial can be found, for example, by use of Schur's formula (Eqs.\
\eqref{eq:Hpartition} and \eqref{eq:schur}). This gives
\begin{equation*}
P(\vec{k},E)=\dtmt{E^2-\mat{H}_{AB}\mat{H}_{AB}^\dagger}.
\end{equation*}
$E^2$, therefore, is an eigenvalue of the $3\times3$ matrix
$\mat{H}_{AB}\mat{H}_{AB}^\dagger$. The latter is readily computed from
\eqnref{eq:Hbipartite}:
\begin{equation*}
\begin{split}
\mat{H}_{AB}\mat{H}_{AB}^\dagger &= \frac{7}{3} +
\begin{pmatrix}
\frac{2}{3} & 1+ e^{-i\theta_2} & e^{i\theta_1}+e^{-i\theta_2} \\
1+ e^{i\theta_2}& -\frac{1}{3} & 1\\
e^{-i\theta_1}+e^{i\theta_2}&1&-\frac{1}{3}
\end{pmatrix} \\
&\equiv \frac{7}{3} + \mat{N}
\end{split}
\end{equation*}
The eigenvalues of the traceless, Hermitian matrix $\mat{N}$ can be
conveniently expressed as
\begin{equation*}
\lambda_n=\rho\cos(\phi+n\frac{2\pi}{3}),\quad n=0,1,2,
\end{equation*}
with
\begin{equation*}
\begin{split}
\rho^2&=\frac{2}{3}\tr(\mat{N}^2)=\frac{8}{3}\left[\frac{8}{3}+
\cos\theta_2+\cos(\theta_1+\theta_2)\right],\\
\cos(3\phi)&=\frac{4}{\rho^3}\dtmt{\mat{N}}\\
&=\frac{4}{\rho^3}\left[\frac{74}{27}+
\frac{8}{3}\Bigl(\cos\theta_2+\cos(\theta_1+\theta_2)\Bigr)
+2\cos\theta_1\right].
\end{split}
\end{equation*}
Now switching to cartesian coordinates of reciprocal space by substituting
$\theta_i=\vec{k}\cdot\vec{a}_i$, we find
\begin{equation*}
\begin{split}
\rho&=\frac{4}{3}\sqrt{4+3\cos\frac{k_x}{2}\cos\frac{\sqrt{3}k_y}{2}},\\
\dtmt{\mat{N}}&=\frac{4}{3}\left[\frac{5}{9}+\cos\frac{k_x}{2}\left(3\cos\frac{k_x}{2}
+4\cos\frac{\sqrt{3}k_y}{2}\right)\right].
\end{split}
\end{equation*}
With these expressions the complete solution for the six band energies is obtained
in closed form: $E=\pm\sqrt{7/3+\lambda_n}$, or
\begin{equation}\label{eq:Eexact}
E=\pm\sqrt{\frac{7}{3}+\rho\cos\left(\frac{1}{3}\arccos\frac{4\dtmt{\mat{N}}}{\rho^3}
+n\frac{2\pi}{3}\right)},\quad
n=0,1,2.
\end{equation}

With the exact solution at hand it is now possible to check that there are indeed two Dirac
points and that they are located at $\vec{k}=\pm(2\pi/3,0)$, as found above. A
Taylor expansion of \eqnref{eq:Eexact} shows that the energy in the
neighborhood of the Dirac points is given by the isotropic dispersion
\begin{equation*}
E=\pm\frac{1}{2\sqrt{3}}\sqrt{\xi^2+\eta^2},
\end{equation*}
where $\xi$ and $\eta$ are the displacement of $k_x$ and $k_y$ from the Dirac
point. This proves the presence of two isotropic Dirac cones. A plot of
\eqnref{eq:Eexact} over the entire Brillouin zone (not shown) reveals no other Dirac
cones.

\begin{figure}
\centering
\includegraphics[width=8.5cm]{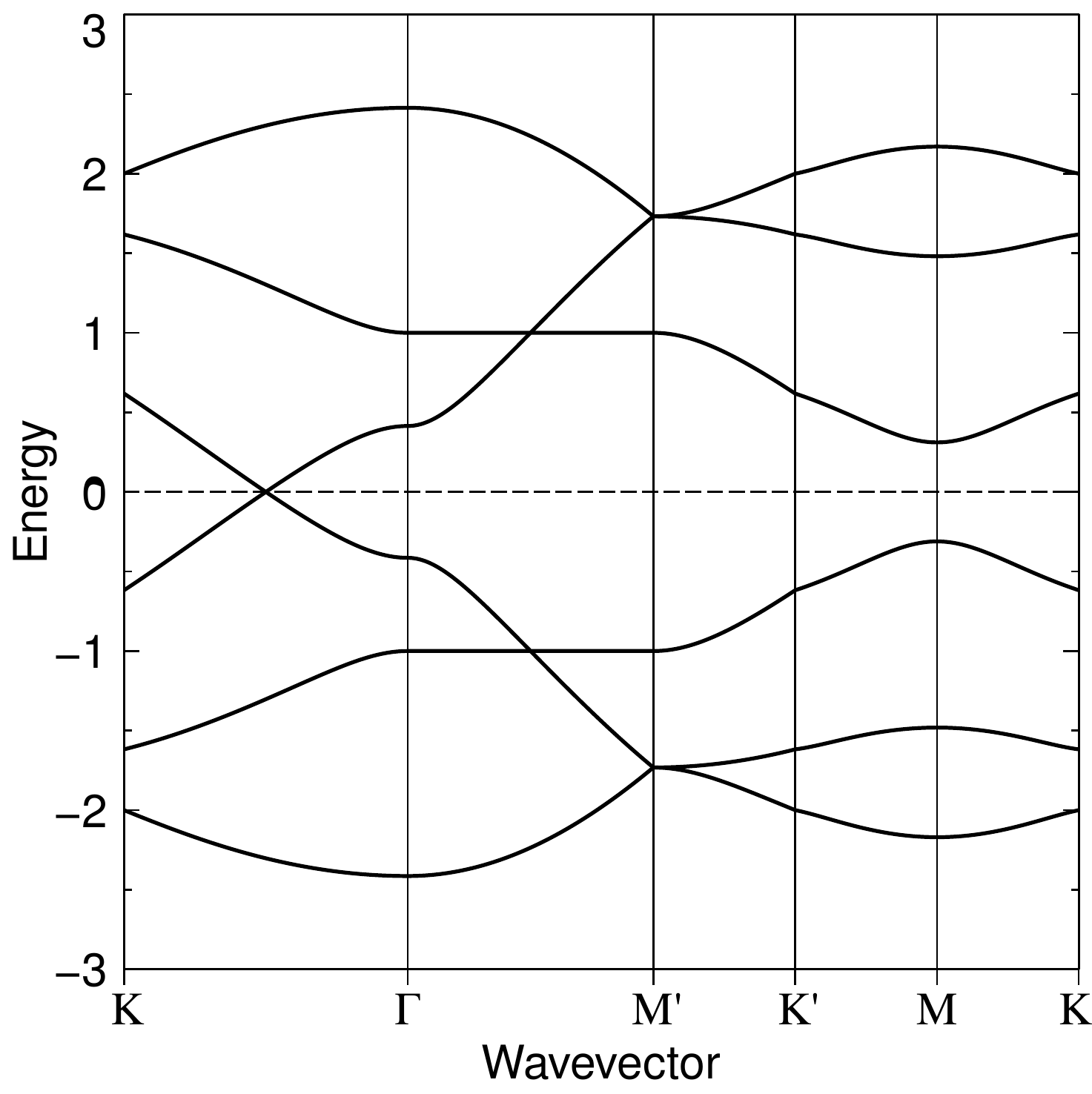}
\caption{The H\"uckel band energies, \eqnref{eq:Eexact}, are plotted along the
boundary of the basic domain, see \figref{fig:BZ}. A Dirac cone appears exactly
halfway between $\Gamma$ and K.}
\label{fig:bands}
\end{figure}

\begin{acknowledgments}
W.V.d.H. is supported by a McKenzie Postdoctoral Fellowship from the University
of Melbourne.
\end{acknowledgments}

\bibliography{dirac_cone}

\begin{thebibliography}{10}%
\makeatletter
\providecommand \@ifxundefined [1]{%
 \ifx #1\undefined \expandafter \@firstoftwo
 \else \expandafter \@secondoftwo
\fi
}%
\providecommand \@ifnum [1]{%
 \ifnum #1\expandafter \@firstoftwo
 \else \expandafter \@secondoftwo
\fi
}%
\providecommand \enquote [1]{``#1''}%
\providecommand \bibnamefont  [1]{#1}%
\providecommand \bibfnamefont [1]{#1}%
\providecommand \citenamefont [1]{#1}%
\providecommand\href[0]{\@sanitize\@href}%
\providecommand\@href[1]{\endgroup\@@startlink{#1}\endgroup\@@href}%
\providecommand\@@href[1]{#1\@@endlink}%
\providecommand \@sanitize [0]{\begingroup\catcode`\&12\catcode`\#12\relax}%
\@ifxundefined \pdfoutput {\@firstoftwo}{%
 \@ifnum{\z@=\pdfoutput}{\@firstoftwo}{\@secondoftwo}%
}{%
 \providecommand\@@startlink[1]{\leavevmode\special{html:<a href="#1">}}%
 \providecommand\@@endlink[0]{\special{html:</a>}}%
}{%
 \providecommand\@@startlink[1]{%
  \leavevmode
  \pdfstartlink
   attr{/Border[0 0 1 ]/H/I/C[0 1 1]}%
   user{/Subtype/Link/A<</Type/Action/S/URI/URI(#1)>>}%
  \relax
 }%
 \providecommand\@@endlink[0]{\pdfendlink}%
}%
\providecommand \url  [0]{\begingroup\@sanitize \@url }%
\providecommand \@url [1]{\endgroup\@href {#1}{\urlprefix}}%
\providecommand \urlprefix [0]{URL }%
\providecommand \Eprint[0]{\href }%
\@ifxundefined \urlstyle {%
  \providecommand \doi [1]{doi:\discretionary{}{}{}#1}%
}{%
  \providecommand \doi [0]{doi:\discretionary{}{}{}\begingroup
  \urlstyle{rm}\Url }%
}%
\providecommand \doibase [0]{http://dx.doi.org/}%
\providecommand \Doi[1]{\href{\doibase#1}}%
\providecommand \selectlanguage [0]{\@gobble}%
\providecommand \bibinfo [0]{\@secondoftwo}%
\providecommand \bibfield [0]{\@secondoftwo}%
\providecommand \translation [1]{[#1]}%
\providecommand \BibitemOpen[0]{}%
\providecommand \bibitemStop [0]{}%
\providecommand \bibitemNoStop [0]{.\EOS\space}%
\providecommand \EOS [0]{\spacefactor3000\relax}%
\providecommand \BibitemShut [1]{\csname bibitem#1\endcsname}%
\bibitem{CastroNeto2009}%
  \BibitemOpen
  \bibfield{author}{%
  \bibinfo {author} {\bibfnamefont{A.~H.}\ \bibnamefont{Castro~Neto}}, \bibinfo
  {author} {\bibfnamefont{F.}~\bibnamefont{Guinea}}, \bibinfo {author}
  {\bibfnamefont{N.~M.~R.}\ \bibnamefont{Peres}}, \bibinfo {author}
  {\bibfnamefont{K.~S.}\ \bibnamefont{Novoselov}},\ and\ \bibinfo {author}
  {\bibfnamefont{A.~K.}\ \bibnamefont{Geim}},\ }%
  \bibfield{journal}{%
  \Doi{10.1103/RevModPhys.81.109}{\bibinfo {journal} {Rev. Mod. Phys.}}\ }%
  \textbf{\bibinfo {volume} {81}},\ \bibinfo {pages} {109} (\bibinfo {year}
  {2009})\BibitemShut{NoStop}%
\bibitem{Novoselov2005}%
  \BibitemOpen
  \bibfield{author}{%
  \bibinfo {author} {\bibfnamefont{K.~S.}\ \bibnamefont{Novoselov}}, \bibinfo
  {author} {\bibfnamefont{A.~K.}\ \bibnamefont{Geim}}, \bibinfo {author}
  {\bibfnamefont{S.~V.}\ \bibnamefont{Morozov}}, \bibinfo {author}
  {\bibfnamefont{D.}~\bibnamefont{Jiang}}, \bibinfo {author}
  {\bibfnamefont{M.~I.}\ \bibnamefont{Katsnelson}}, \bibinfo {author}
  {\bibfnamefont{I.~V.}\ \bibnamefont{Grigorieva}}, \bibinfo {author}
  {\bibfnamefont{S.~V.}\ \bibnamefont{Dubonos}},\ and\ \bibinfo {author}
  {\bibfnamefont{A.~A.}\ \bibnamefont{Firsov}},\ }%
  \bibfield{journal}{%
  \bibinfo {journal} {Nature}\ }%
  \textbf{\bibinfo {volume} {438}},\ \bibinfo {pages} {197} (\bibinfo {year}
  {2005})\BibitemShut{NoStop}%
\bibitem{Katsnelson2006}%
  \BibitemOpen
  \bibfield{author}{%
  \bibinfo {author} {\bibfnamefont{M.~I.}\ \bibnamefont{Katsnelson}}, \bibinfo
  {author} {\bibfnamefont{K.~S.}\ \bibnamefont{Novoselov}},\ and\ \bibinfo
  {author} {\bibfnamefont{A.~K.}\ \bibnamefont{Geim}},\ }%
  \bibfield{journal}{%
  \bibinfo {journal} {Nature Phys.}\ }%
  \textbf{\bibinfo {volume} {2}},\ \bibinfo {pages} {620} (\bibinfo {year}
  {2006})\BibitemShut{NoStop}%
\bibitem{Kim2012}%
  \BibitemOpen
  \bibfield{author}{%
  \bibinfo {author} {\bibfnamefont{B.~G.}\ \bibnamefont{Kim}}\ and\ \bibinfo
  {author} {\bibfnamefont{H.~J.}\ \bibnamefont{Choi}},\ }%
  \bibfield{journal}{%
  \Doi{10.1103/PhysRevB.86.115435}{\bibinfo {journal} {Phys. Rev. B}}\ }%
  \textbf{\bibinfo {volume} {86}},\ \bibinfo {pages} {115435} (\bibinfo {year}
  {2012})\BibitemShut{NoStop}%
\bibitem{Zheng2013}%
  \BibitemOpen
  \bibfield{author}{%
  \bibinfo {author} {\bibfnamefont{J.-J.}\ \bibnamefont{Zheng}}, \bibinfo
  {author} {\bibfnamefont{X.}~\bibnamefont{Zhao}}, \bibinfo {author}
  {\bibfnamefont{S.~B.}\ \bibnamefont{Zhang}},\ and\ \bibinfo {author}
  {\bibfnamefont{X.}~\bibnamefont{Gao}},\ }%
  \bibfield{journal}{%
  \Doi{http://dx.doi.org/10.1063/1.4811841}{\bibinfo {journal} {J. Chem.
  Phys.}}\ }%
  \textbf{\bibinfo {volume} {138}},\ \bibinfo {eid} {244708} (\bibinfo {year}
  {2013})\BibitemShut{NoStop}%
\bibitem{Ouyang2011}%
  \BibitemOpen
  \bibfield{author}{%
  \bibinfo {author} {\bibfnamefont{F.}~\bibnamefont{Ouyang}}, \bibinfo {author}
  {\bibfnamefont{S.}~\bibnamefont{Peng}}, \bibinfo {author}
  {\bibfnamefont{Z.}~\bibnamefont{Liu}},\ and\ \bibinfo {author}
  {\bibfnamefont{Z.}~\bibnamefont{Liu}},\ }%
  \bibfield{journal}{%
  \Doi{10.1021/nn200580w}{\bibinfo {journal} {ACS Nano}}\ }%
  \textbf{\bibinfo {volume} {5}},\ \bibinfo {pages} {4023} (\bibinfo {year}
  {2011})\BibitemShut{NoStop}%
\bibitem{Yang2012}%
  \BibitemOpen
  \bibfield{author}{%
  \bibinfo {author} {\bibfnamefont{M.}~\bibnamefont{Yang}}, \bibinfo {author}
  {\bibfnamefont{L.}~\bibnamefont{Zhou}}, \bibinfo {author}
  {\bibfnamefont{J.}~\bibnamefont{Wang}}, \bibinfo {author}
  {\bibfnamefont{Z.}~\bibnamefont{Liu}},\ and\ \bibinfo {author}
  {\bibfnamefont{Z.}~\bibnamefont{Liu}},\ }%
  \bibfield{journal}{%
  \Doi{10.1021/jp2088143}{\bibinfo {journal} {J. Phys. Chem. C}}\ }%
  \textbf{\bibinfo {volume} {116}},\ \bibinfo {pages} {844} (\bibinfo {year}
  {2012})\BibitemShut{NoStop}%
\bibitem{Zhang2013}%
  \BibitemOpen
  \bibfield{author}{%
  \bibinfo {author} {\bibfnamefont{L.}~\bibnamefont{Zhang}}, \bibinfo {author}
  {\bibfnamefont{J.}~\bibnamefont{Yu}}, \bibinfo {author}
  {\bibfnamefont{M.}~\bibnamefont{Yang}}, \bibinfo {author}
  {\bibfnamefont{Q.}~\bibnamefont{Xie}}, \bibinfo {author}
  {\bibfnamefont{H.}~\bibnamefont{Peng}},\ and\ \bibinfo {author}
  {\bibfnamefont{Z.}~\bibnamefont{Liu}},\ }%
  \bibfield{journal}{%
  \bibinfo {journal} {Nat. Commun.}\ }%
  \textbf{\bibinfo {volume} {4}},\ \bibinfo {pages} {1443} (\bibinfo {year}
  {2013})\BibitemShut{NoStop}%
\bibitem{Ma2013}%
  \BibitemOpen
  \bibfield{author}{%
  \bibinfo {author} {\bibfnamefont{Y.}~\bibnamefont{Ma}}, \bibinfo {author}
  {\bibfnamefont{Y.}~\bibnamefont{Dai}},\ and\ \bibinfo {author}
  {\bibfnamefont{B.}~\bibnamefont{Huang}},\ }%
  \bibfield{journal}{%
  \Doi{10.1021/jz401099j}{\bibinfo {journal} {J. Phys. Chem. Lett.}}\ }%
  \textbf{\bibinfo {volume} {4}},\ \bibinfo {pages} {2471} (\bibinfo {year}
  {2013})\BibitemShut{NoStop}%
\bibitem{Soncini2001}%
  \BibitemOpen
  \bibfield{author}{%
  \bibinfo {author} {\bibfnamefont{A.}~\bibnamefont{Soncini}}, \bibinfo
  {author} {\bibfnamefont{P.~W.}\ \bibnamefont{Fowler}}, \bibinfo {author}
  {\bibfnamefont{I.}~\bibnamefont{Cernusak}},\ and\ \bibinfo {author}
  {\bibfnamefont{E.}~\bibnamefont{Steiner}},\ }%
  \bibfield{journal}{%
  \Doi{10.1039/B103929F}{\bibinfo {journal} {Phys. Chem. Chem. Phys.}}\ }%
  \textbf{\bibinfo {volume} {3}},\ \bibinfo {pages} {3920} (\bibinfo {year}
  {2001})\BibitemShut{NoStop}%
\bibitem{Soncini2005a}%
  \BibitemOpen
  \bibfield{author}{%
  \bibinfo {author} {\bibfnamefont{A.}~\bibnamefont{Soncini}}, \bibinfo
  {author} {\bibfnamefont{P.~W.}\ \bibnamefont{Fowler}},\ and\ \bibinfo
  {author} {\bibfnamefont{L.~W.}\ \bibnamefont{Jenneskens}},\ }%
  \bibfield{journal}{%
  \bibinfo {journal} {Struc. Bond.}\ }%
  \textbf{\bibinfo {volume} {115}},\ \bibinfo {pages} {57} (\bibinfo {year}
  {2005})\BibitemShut{NoStop}%
\bibitem{Soncini2005}%
  \BibitemOpen
  \bibfield{author}{%
  \bibinfo {author} {\bibfnamefont{A.}~\bibnamefont{Soncini}}, \bibinfo
  {author} {\bibfnamefont{C.}~\bibnamefont{Domene}}, \bibinfo {author}
  {\bibfnamefont{J.~J.}\ \bibnamefont{Engelberts}}, \bibinfo {author}
  {\bibfnamefont{P.~W.}\ \bibnamefont{Fowler}}, \bibinfo {author}
  {\bibfnamefont{A.}~\bibnamefont{Rassat}}, \bibinfo {author}
  {\bibfnamefont{J.~H.}\ \bibnamefont{van Lenthe}}, \bibinfo {author}
  {\bibfnamefont{R.~W.~A.}\ \bibnamefont{Havenith}},\ and\ \bibinfo {author}
  {\bibfnamefont{L.~W.}\ \bibnamefont{Jenneskens}},\ }%
  \bibfield{journal}{%
  \Doi{10.1002/chem.200400678}{\bibinfo {journal} {Chem. Eur. J.}}\ }%
  \textbf{\bibinfo {volume} {11}},\ \bibinfo {pages} {1257} (\bibinfo {year}
  {2005})\BibitemShut{NoStop}%
\bibitem{Fowler2001}%
  \BibitemOpen
  \bibfield{author}{%
  \bibinfo {author} {\bibfnamefont{P.~W.}\ \bibnamefont{Fowler}}, \bibinfo
  {author} {\bibfnamefont{R.~W.~A.}\ \bibnamefont{Havenith}}, \bibinfo {author}
  {\bibfnamefont{L.~W.}\ \bibnamefont{Jenneskens}}, \bibinfo {author}
  {\bibfnamefont{A.}~\bibnamefont{Soncini}},\ and\ \bibinfo {author}
  {\bibfnamefont{E.}~\bibnamefont{Steiner}},\ }%
  \bibfield{journal}{%
  \Doi{10.1039/B106651J}{\bibinfo {journal} {Chem. Commun.}},\ \bibinfo {pages}
  {2386}}%
   (\bibinfo {year} {2001})\BibitemShut{NoStop}%
\bibitem{Soncini2002}%
  \BibitemOpen
  \bibfield{author}{%
  \bibinfo {author} {\bibfnamefont{A.}~\bibnamefont{Soncini}}, \bibinfo
  {author} {\bibfnamefont{R.~W.~A.}\ \bibnamefont{Havenith}}, \bibinfo {author}
  {\bibfnamefont{P.~W.}\ \bibnamefont{Fowler}}, \bibinfo {author}
  {\bibfnamefont{L.~W.}\ \bibnamefont{Jenneskens}},\ and\ \bibinfo {author}
  {\bibfnamefont{E.}~\bibnamefont{Steiner}},\ }%
  \bibfield{journal}{%
  \Doi{10.1021/jo020091d}{\bibinfo {journal} {J. Org. Chem.}}\ }%
  \textbf{\bibinfo {volume} {67}},\ \bibinfo {pages} {4753} (\bibinfo {year}
  {2002})\BibitemShut{NoStop}%
\bibitem{Horn1990}%
  \BibitemOpen
  \bibfield{author}{%
  \bibinfo {author} {\bibfnamefont{R.~A.}\ \bibnamefont{Horn}}\ and\ \bibinfo
  {author} {\bibfnamefont{C.~R.}\ \bibnamefont{Johnson}},\ }%
  \emph{\bibinfo {title} {Matrix analysis}}\ (\bibinfo {publisher} {Cambridge
  University Press},\ \bibinfo {year} {1990})\BibitemShut{NoStop}%
\bibitem{Note1}%
  \BibitemOpen
  \bibinfo {note} {Assuming for now that all three bonds are decorated, so that
  $\gamma =0$. This is done for generality only; it does not represent a
  limitation of the theory. The example in Sec.\ \ref {sec:example} will show
  how to treat undecorated bonds on the same footing as decorated
  bonds.}\BibitemShut{Stop}%
\bibitem{Pickup2008}%
  \BibitemOpen
  \bibfield{author}{%
  \bibinfo {author} {\bibfnamefont{B.~T.}\ \bibnamefont{Pickup}}\ and\ \bibinfo
  {author} {\bibfnamefont{P.~W.}\ \bibnamefont{Fowler}},\ }%
  \bibfield{journal}{%
  \Doi{http://dx.doi.org/10.1016/j.cplett.2008.05.062}{\bibinfo {journal}
  {Chem. Phys. Lett.}}\ }%
  \textbf{\bibinfo {volume} {459}},\ \bibinfo {pages} {198 } (\bibinfo {year}
  {2008})\BibitemShut{NoStop}%
\bibitem{Gutman1986}%
  \BibitemOpen
  \bibfield{author}{%
  \bibinfo {author} {\bibfnamefont{I.}~\bibnamefont{Gutman}}\ and\ \bibinfo
  {author} {\bibfnamefont{O.~E.}\ \bibnamefont{Polansky}},\ }%
  \emph{\bibinfo {title} {Mathematical concepts in organic chemistry}}\
  (\bibinfo {publisher} {Springer},\ \bibinfo {address} {Berlin},\ \bibinfo
  {year} {1986})\BibitemShut{NoStop}%
\bibitem{Hasegawa2006}%
  \BibitemOpen
  \bibfield{author}{%
  \bibinfo {author} {\bibfnamefont{Y.}~\bibnamefont{Hasegawa}}, \bibinfo
  {author} {\bibfnamefont{R.}~\bibnamefont{Konno}}, \bibinfo {author}
  {\bibfnamefont{H.}~\bibnamefont{Nakano}},\ and\ \bibinfo {author}
  {\bibfnamefont{M.}~\bibnamefont{Kohmoto}},\ }%
  \bibfield{journal}{%
  \Doi{10.1103/PhysRevB.74.033413}{\bibinfo {journal} {Phys. Rev. B}}\ }%
  \textbf{\bibinfo {volume} {74}},\ \bibinfo {pages} {033413} (\bibinfo {year}
  {2006})\BibitemShut{NoStop}%
\bibitem{Coulson1940}%
  \BibitemOpen
  \bibfield{author}{%
  \bibinfo {author} {\bibfnamefont{C.~A.}\ \bibnamefont{Coulson}}\ and\
  \bibinfo {author} {\bibfnamefont{G.~S.}\ \bibnamefont{Rushbrooke}},\ }%
  \bibfield{journal}{%
  \bibinfo {journal} {Proc. Camb. Phil. Soc.}\ }%
  \textbf{\bibinfo {volume} {36}},\ \bibinfo {pages} {193} (\bibinfo {year}
  {1940})\BibitemShut{NoStop}%
\end{thebibliography}%

\end{document}